\documentclass[twocolumn]{aastex631}
\usepackage{apjfonts}
\usepackage{graphics,graphicx,float,import,enumitem,booktabs}

\newcommand\hii{\ion{H}{2} }

\newcommand\oi{[\ion{O}{1}]}
\newcommand\oii{[\ion{O}{2}]}
\newcommand\oiii{[\ion{O}{3}]}
\newcommand\neiii{[\ion{Ne}{3}]}
\newcommand\siii{[\ion{S}{3}]}
\newcommand\nii{[\ion{N}{2}]}
\newcommand\sii{[\ion{S}{2}]}
\newcommand\ariii{[\ion{Ar}{3}]}
\newcommand\ariv{[\ion{Ar}{4}]}
\newcommand\te{T$_e$}
\newcommand\den{n$_e$}

\shorttitle{Direct Gas-Phase Abundances at $z=2.96$}
\shortauthors{Rogers et al.}
\accepted{to ApJL on March 1st, 2024}

\begin{document}

\title{CECILIA: Direct O, N, S, and Ar Abundances in Q2343-D40, a Galaxy at $z\sim$3}

\author[0000-0002-0361-8223]{Noah S. J. Rogers}
\affiliation{Center for Interdisciplinary Exploration and Research in Astrophysics (CIERA), Northwestern University, 1800 Sherman Ave., Evanston, IL, 60201, USA}

\author[0000-0001-6369-1636]{Allison L. Strom}
\affiliation{Center for Interdisciplinary Exploration and Research in Astrophysics (CIERA), Northwestern University, 1800 Sherman Ave., Evanston, IL, 60201, USA}
\affiliation{Department of Physics and Astronomy, Northwestern University, 2145 Sheridan Road, Evanston, IL 60208, USA}

\author[0000-0002-8459-5413]{Gwen C. Rudie}
\affiliation{Carnegie Observatories, 813 Santa Barbara Street, Pasadena, CA 91101, USA}

\author[0000-0002-6967-7322]{Ryan F. Trainor}
\affiliation{Department of Physics and Astronomy, Franklin \& Marshall College, 637 College Avenue, Lancaster, PA 17603, USA}

\author[0009-0008-2226-5241]{Menelaos Raptis}
\affiliation{Department of Physics and Astronomy, Franklin \& Marshall College, 637 College Avenue, Lancaster, PA 17603, USA}

\author[0000-0002-6034-082X]{Caroline von Raesfeld}
\affiliation{Center for Interdisciplinary Exploration and Research in Astrophysics (CIERA), Northwestern University, 1800 Sherman Ave., Evanston, IL, 60201, USA}
\affiliation{Department of Physics and Astronomy, Northwestern University, 2145 Sheridan Road, Evanston, IL 60208, USA}

\begin{abstract}

Measurements of chemical abundances in high-$z$ star-forming (SF) galaxies place important constraints on the enrichment histories of galaxies and the physical conditions in the early universe. JWST is beginning to enable direct chemical abundance measurements in galaxies at $z$$>$2 via the detection of the faint \te-sensitive auroral line \oiii$\lambda$4364. However, abundances of other elements (e.g., S and Ar) in high-$z$ galaxies remain unconstrained due to a lack of \te\ data and wavelength coverage. Here, we present multiple direct abundances in Q2343-D40, a galaxy at $z=$2.9628$\pm$0.0001 observed with JWST/NIRSpec as part of the CECILIA program. We report the first simultaneous measurement of \te\oiii\ and \te\siii\ in a high-$z$ galaxy, finding good agreement with the temperature trends in local SF systems. We measure a gas-phase metallicity of 12+log(O/H) $=8.07\pm0.06$, and the N/O abundance, log(N/O) $=-1.37\pm0.21$, is indicative of primary nucleosynthesis. The S/O and Ar/O relative abundances, log(S/O)$=-1.88\pm0.10$ and log(Ar/O)$=-2.80\pm0.12$, are both $>$0.3 dex lower than the solar ratios. However, the relative Ar$^{2+}$/S$^{2+}$ abundance is consistent with the solar ratio, suggesting that the relative S-to-Ar abundance does not evolve significantly with redshift. Recent nucleosynthesis models find that a significant amount of S and Ar are produced in Type Ia supernovae, such that the S/O and Ar/O abundances in Q2343-D40 could be the result of predominantly core-collapse supernovae enrichment. Future JWST observations of high-$z$ galaxies will uncover whether S/O and Ar/O are sensitive to the timescales of these different enrichment mechanisms.

\end{abstract}

\section{Introduction}

The buildup of heavy elements is a direct consequence of galaxy evolution. Episodes of star formation produce massive stars, which in turn synthesize heavy metals through stellar nucleosynthesis. By means of mass-loss events and supernovae, the products of stellar nucleosynthesis are returned to the interstellar medium (ISM). These metals mix with the ISM and accreting gas, resulting in an increased gas-phase abundance that is imprinted in the next generation of stars. Therefore, the gas-phase metal abundance is sensitive to the past star formation, gas inflows and outflows, and ISM mixing mechanisms within a galaxy \citep[e.g.,][]{tosi1988,roy1995,berg2019}. Chemical abundance trends are of particular importance at high-$z$, as they provide insight into the gas-phase conditions in the early universe that give rise to the galaxy scaling relations observed today \citep[such as the mass-metallicity relation,][]{lequ1979,trem2004,erb2006}.

Additionally, the relative abundances of different elements provides insight into the star-formation history of a galaxy and nucleosynthetic mechanisms that operate on different timescales. For instance, O is produced in massive stars and returned to the ISM on short timescales via core-collapse supernovae (CCSN), such that gas-phase O/H traces the enrichment from massive stars. Massive stars also synthesize elements like Ne, S, and Ar through the $\alpha$ process. If this is the principal process by which these elements are produced, then the relative $\alpha$ element abundances (e.g., Ne/O) will be independent of the overall gas-phase metallicity (i.e, O/H). While N is synthesized in massive stars, N also has a secondary, metallicity-sensitive enrichment mechanism from intermediate-mass stars \citep{henr2000}. Consequently, the N/O relative abundance is dependent on both the enrichment from stars of different masses and the overall metallicity of the ISM. Finally, Fe is produced in CCSN and Type Ia SNe; while Type Ia SNe contribute significantly to the overall Fe abundance, the timescale required for Type Ia enrichment is much longer than CCSN. This combination makes Fe/H sensitive to the prior star-formation history, and the $\alpha$/Fe ratio is dependent on the relative enrichment of CCSN and Type Ia SNe.

Reliable chemical abundances can be measured from the collisionally-excited lines (CELs) of metal ions in the ISM. The relative intensity of two CELs originating from different energy levels in the same metal ion is sensitive to gas-phase physical conditions such as the electron gas temperature \te\ and electron density \den. With these conditions, ionic abundances in the ISM can be directly calculated from the intensity and emissivity of metal CELs. While the intense ``nebular'' CELs can consistently be observed even in high-$z$ nebulae \citep[e.g.,][]{bunk2023,naka2023,will2023}, the ``direct'' abundance method \citep{dine1990} relies on the simultaneous measurement of these CELs and faint, \te-sensitive auroral lines in the rest-frame optical. Despite observational challenges associated with this technique, including potential biases in the presence of \te\ and \den\ variations \citep[e.g.,][]{peim1967,rubi1989,mend2023,mend2023N}, the direct method continues to be an important and successful tool for understanding the chemical enrichment of the local universe \citep[e.g.,][among many others]{kenn2003b,este2004,bres2011,berg2015,arel2020M,roge2022}.

Until recently, the direct abundance method was difficult to apply in distant sources. At $z$$\sim$1, the strong \oiii\ CELs redshift into the near-infrared (NIR) and the \oiii\ auroral line becomes exceedingly faint in low surface brightness sources. The same is true for the rest-frame NIR lines of \siii\ and \ariii: the highest redshift measurements of direct S$^{2+}$/H$^+$ and Ar$^{2+}$/H$^+$ using these emission lines are in ionized regions at $z$$\lesssim$0.3 \citep[e.g.,][]{izot2021} and $z$$\le$0.1 \citep{dors2023}, respectively. While measurements of \te\oiii\ and direct O/H abundances have been acquired in individual galaxies at $z$$>$1 through transmission windows in the NIR \citep[see samples in][]{sand2020,clar2023} and from UV spectroscopy of highly lensed systems \citep[e.g.,][]{jame2014,citr2023}, statistically significant samples of direct O/H abundances (let alone abundances from other elements such as N and S) have been challenging to acquire. 

However, with the NIR capabilities of the \textit{James Webb Space Telescope} (JWST), direct \te\ measurements and abundances are now readily accessible in high-$z$ galaxies \citep[e.g.,][]{arel2022,scha2022,trum2023,sand2023,lase2023,welc2024,topp2024}. While these studies have started to uncover the chemical abundance patterns in the early universe, the current scope of high-$z$ chemical abundances is still fairly limited. First, it is often the case that only a single auroral line, \oiii$\lambda$4364, is available in the spectra of high-$z$ star-forming galaxies. This auroral line permits a direct measurement of \te\oiii, which primarily traces the gas containing high-ionization ions such as O$^{2+}$ and Ne$^{2+}$. However, a lack of \te\ in other ionization zones can significantly bias direct abundance measurements \citep{arel2020,roge2022}. Particularly, robust S$^{2+}$ and Ar$^{2+}$ abundances require a measurement of \te\siii, which has not been made in a galaxy at $z$$>$0.2. Second, these prior surveys have focused mainly on the gas-phase O/H abundance and the relative Ne/O abundance if \neiii$\lambda$3870 is detected. Measuring the abundance of other elements (e.g., N, S, and Ar) in high-$z$ galaxies requires both deep spectroscopy and broad NIR wavelength coverage.

The CECILIA program \citep{stro2023} is poised to contribute detailed measurements of high-$z$ \te\ and ionic abundances from numerous ions. CECILIA uses the G235M and G395M gratings of NIRSpec and targets star-forming galaxies at $z$$>$2 for ultra-deep NIR spectroscopy, enabling the detection of the \siii$\lambda$6314 and \oii$\lambda\lambda$7322,7332 auroral lines and strong NIR lines of \nii, \sii, \siii, and \ariii. Within this sample, the galaxy Q2343-D40 is of particular interest. Owing to Q2343-D40's spectroscopic redshift of $z$$\sim$3, simultaneous measurements of \te\oiii, \te\siii, and robust ionic abundances are possible from the NIRSpec data. Here, we report the physical conditions (\te\oiii, \te\siii, and \den\sii) and chemical abundances of O, N, S, and Ar in Q2343-D40, a star-forming galaxy $\sim$2 Gyr after the Big Bang. This manuscript is organized as follows: in \S2 we briefly summarize the data reduction, emission line fitting, and reddening corrections applied to the Q2343-D40 spectrum; the \te, \den, ionic abundance, ionization correction factors (ICFs), and total abundance calculations are described in \S3; we compare Q2343-D40's direct abundance trends to local star-forming systems and discuss these findings in \S4; we summarize our conclusions in \S5. Although all reported abundances are \te-based, direct abundances refer to those that are derived from a measured \te, while inferred abundances are those determined through application of a \te-\te\ relation or ICF. In this work, we assume a $\Lambda$CDM cosmology with $H_0$$=$70 km s$^{-1}$ Mpc$^{-1}$, $\Omega_m$$=$0.3, and $\Omega_\Lambda$$=$0.7. We adopt the solar abundance values from \citet{aspl2021}: 12+log(O/H)$_\odot = 8.69\pm0.04$ dex, log(N/O)$_\odot = -0.86\pm0.08$ dex, log(S/O)$_\odot = -1.57\pm0.05$ dex, and log(Ar/O)$_\odot = -2.37\pm0.11$ dex. Throughout the text, we refer to emission lines using their vacuum wavelengths.

\section{Observations and Reduction}

Q2343-D40 (R.A.$=$23h46m22.8s, Decl.$=$12$^{\circ}$49\arcmin32\farcs1), hereafter D40, was observed as part of the CECILIA program, which targeted rest-UV color-selected star-forming galaxies in the Keck Baryonic Structure Survey \citep[KBSS,][]{stei2010,rudi2012,trai2015,stro2017}. We briefly highlight the most important aspects of the observations here and refer the reader to \citet{stro2023} for the details concerning sample selection, exposure time requirements, and data reduction. Two NIRSpec grating combinations were used: G235M ($R$$\sim$1000, 1.66$-$3.07$\mu$m) data were acquired with a 29.5 hour exposure time to reliably detect the \siii$\lambda$6314 and \oii$\lambda$7322,7332 auroral lines; G395M ($R$$\sim$1000, 2.87$-$5.10$\mu$m) data, necessary to detect \siii$\lambda\lambda$9071,9533 for targets at $z$$\gtrsim$2.4, were taken with an exposure time of 1.1 hours.

While the G235M data reduction was described in \citet{stro2023}, there have been significant improvements to the calibration files in recent versions of the \textsc{calwebb} pipeline, including updated reference pixels and bar shadow reference file for the MSA data. We reran the \textsc{calwebb} pipeline using version 1.12.5 \citep[\textsc{crds\textunderscore context} $=$ jwst\textunderscore1180.pmap,][]{calwebb_v1.12.5_2023}, \textsc{msaexp} version 0.6.11 \citep{msaexp2022}, and \textsc{grizli} version 1.8.9 \citep{griz2023}. We use \textsc{NSClean} \citep{raus2023} to correct for $1/f$ noise. However, we observed that running \textsc{NSClean} with default parameters introduced residual noise artifacts in the G235M and G395M rate files. Updating the critical frequency to fc$=$1/2048, the kill width to kw$=$fc/4, and the Gaussian smoothing standard deviation to buffer\textunderscore sigma$=$3 removed the $1/f$ noise while decreasing the residual RMS noise in the individual rate files.

Background subtraction was completed using a global background solution based on all available MSA slits (masking pixels illuminated by the science targets) for both G235M and G395M data \citep[similar to][]{stro2023}. To ensure reliable extractions of the science spectra, we fit and subtract the residual background in D40's 2D spectra with a low-order polynomial in the dispersion direction. 1D science spectra are extracted using the default \textsc{msaexp} Gaussian extraction model. Despite these updates, there are persistent flux calibration issues in the NIRSpec data. To account for these, we rescale the G235M spectrum to match the best-fit model spectral energy distribution (SED) of D40. The SED fit is based on existing optical (\textit{U$_n$}, \textit{G}, and $\mathcal{R}$) and NIR (\textit{H} and \textit{K$_s$}) photometry, as well as imaging using \textit{Hubble Space Telescope}/WFC3 F140W. We calculate the ratio of the SED and NIRSpec spectrum, then fit this ratio with a quadratic polynomial as a function of wavelength while excluding emission lines. This flux rescaling function, $f_{corr}(\lambda)$, is used to scale the G235M spectrum to the SED continuum \citep[similar to the approach taken by][]{arra2023}.

\begin{figure*}[!t]
   \centering
   \includegraphics[width=0.50\textwidth]{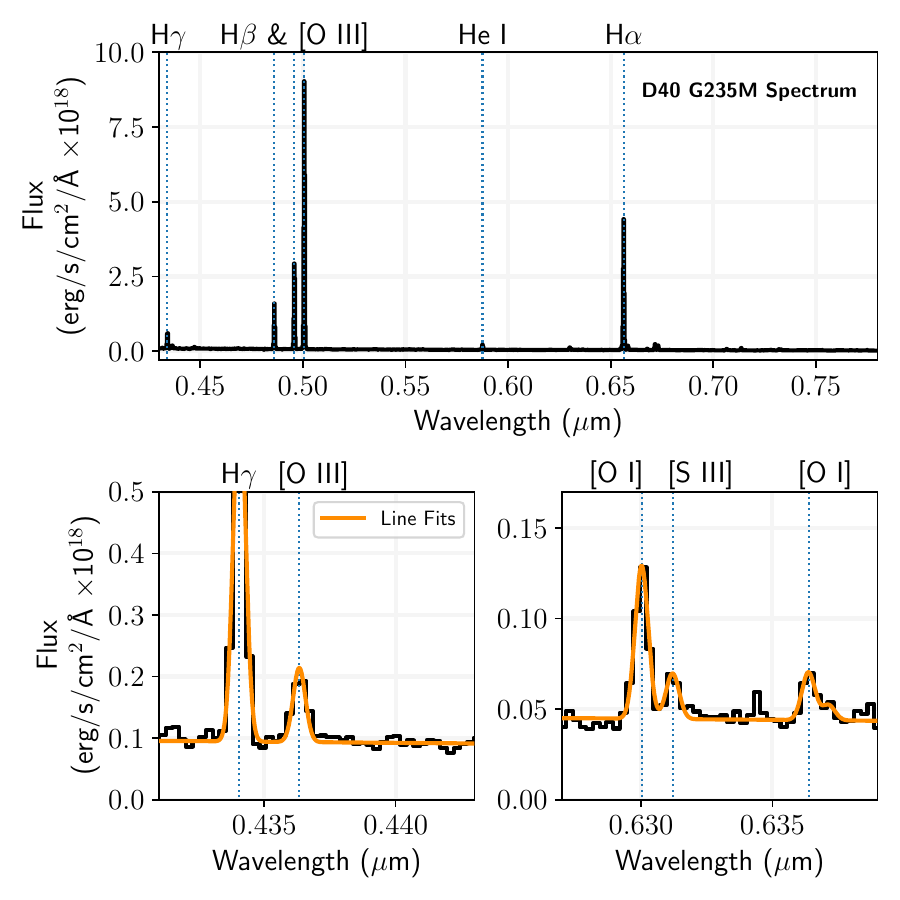}
   \hskip -2ex
   \includegraphics[width=0.50\textwidth]{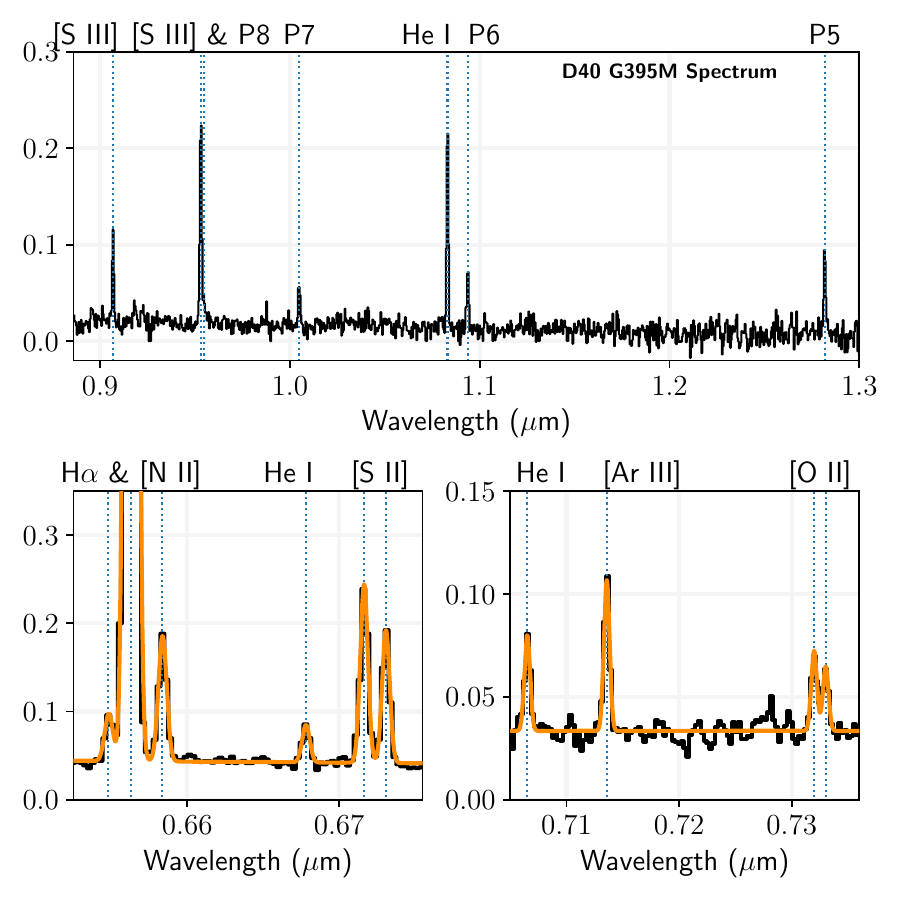}
   \caption{The redshift-corrected NIRSpec spectrum of D40. The top two panels provide the full spectral range of the G235M (left) and G395M (right) spectra, while panels in the bottom row highlight the faint emission lines detected in the G235M spectrum. In each panel, key spectral features are denoted with vertical dotted lines. The Gaussian+linear continuum fits to the emission lines are shown in orange. The auroral lines \oiii$\lambda$4364 and \siii$\lambda$6314 are detected at high significance, producing the data required to measure, for the first time in a high-$z$ galaxy, simultaneous \te\oiii\ and \te\siii\ and the gas-phase abundances from N, O, S, and Ar.}
   \label{fig:spectrum}
\end{figure*} 

Estimating a similar flux rescaling function for the G395M data is challenging owing to the weaker continuum with higher RMS noise. Instead, we use the spectral region that is coincident between G235M and G395M to scale the G395M continuum to the G235M continuum. We note that these flux differences between the two gratings could be due to a combination of flux calibration issues and the wavelength-dependent point spread function. For the following analysis, we use line flux ratios within individual filters to avoid potential systematic errors associated with the cross-band flux calibration. The redshift of D40 was uncertain prior to the NIRSpec observations, but we determine the spectroscopic redshift of D40 to be $z$$=$2.9628$\pm$0.0001 using 11 strong emission lines.

The final reduced rest-frame spectrum of D40 is plotted in Figure \ref{fig:spectrum}, where the strong nebular lines are highlighted by vertical dotted lines in each panel. Emission lines are fit using a linear continuum and Gaussian profile: the G235M (G395M) spectrum is split into 9 (7) windows, where a constant full-width at half maximum (FWHM) and velocity offset are used for all lines in a given window. This procedure allows for a fit to emission lines that are blended at NIRSpec's resolution (e.g., [\ion{Ar}{4}]$\lambda$4713 and \ion{He}{1}$\lambda$4715). To account for stellar absorption, the Balmer lines are fit with Gaussian profiles and the best-fit SED continuum, which shows relatively low Balmer absorption equivalent width ($\lesssim$3.5 \AA). We repeated the analysis using the SED continuum for all emission line fits and find that the results are consistent with the linear continuum+Gaussian emission line approach. Therefore, we adopt the linear continuum+Gaussian line model for all other emission lines to allow for a more flexible fit to the local continuum. We measure the RMS noise in continuum windows on either side of the emission line, then calculate the line flux uncertainty as $\delta F(\lambda) = 2\times\sqrt{n_{pix}}\times RMS$, where $n_{pix}$ is the number of pixels within $\pm$1 FWHM from the line center \citep[see][]{berg2013}. Although this does not account for the uncertainty in the absolute flux calibration, the analysis presented here focuses on the ratios of line fluxes.

To estimate the relative flux uncertainties, we utilize the G235M $f_{corr}(\lambda)$ functions for the 23 CECILIA galaxies reported in \citet{stro2023}. We normalize each galaxy's $f_{corr}(\lambda)$ function at the wavelength of H$\beta$ measured in D40, or 1.93$\mu$m. For the full sample, we calculate the standard deviation in $f_{corr}(\lambda)$/$f_{corr}(1.93\mu$m$)$ as a function of wavelength. We find that the standard deviation measured from the ensemble of $f_{corr}(\lambda)$/$f_{corr}(1.93\mu$m$)$ is relatively constant and $\lesssim$6\%\ at a given wavelength. As such, we assume that a constant 6\% uncertainty on all flux ratios relative to H$\beta$ provides a reasonable approximation on the relative flux calibration uncertainty. This uncertainty is added in quadrature to the uncertainty on all emission line flux ratios before calculating physical conditions. An emission line is considered detected if $F(\lambda)$/$F($H$\beta)$ has a signal-to-noise (S/N) $>$ 3.

To correct for dust attenuation, we first account for attenuation along the line of sight to D40. The foreground reddening, E(B$-$V)$=$0.037, was retrieved from the \citet{schl1998} dust maps using the \textsc{Python dustmaps} package \citep{gree2018} and the position of D40. The observed spectrum is corrected for foreground extinction using the \citet{card1989} Galactic attenuation law with $R_V$$=$3.1. Although the \citet{card1989} is only calibrated to 3.3$\mu$m, the low foreground attenuation makes this correction relatively small for D40 (i.e., the maximum correction within the G235M spectrum is a $\sim$2\% increase in the flux of H$\gamma$). We use the \textsc{Python PyNeb} \citep{luri2012,luri2015L} package's \textsc{getCorr} function to perform the correction for foreground attenuation.

Next, we determine E(B$-$V) due to dust attenuation within D40 using the \citet{redd2020} quadratic nebular attenuation law. This law has been calibrated on the Balmer line emission in star-forming galaxies at $1.4\leq z \leq2.6$ from the MOSFIRE Deep Evolution Field survey \citep{krie2015}, which should better characterize the general trends of dust attenuation for high-$z$ galaxies. The shape of the \citet{redd2020} law is similar to that of the \citet{card1989} Galactic attenuation law and implies an $R_V$ of 3.09. Future JWST spectroscopic observations will better constrain the shape of the dust attenuation law at high-$z$ and will assess whether its functional form deviates significantly from the frequently utilized Galactic attenuation law \citep[see recent work by][]{mark2024}.

The best-fit E(B$-$V) is calculated as a weighted average from the E(B$-$V) predicted by the observed H$\alpha$/H$\beta$ and H$\gamma$/H$\beta$ ratios. Both Balmer line ratios are compared to their predicted intensity ratios from \citet{stor1995} assuming Case B recombination at \te=1.2$\times$10$^4$ K and \den$=$10$^2$ cm$^{-3}$. For each ratio, uncertainty is estimated by recalculating E(B$-$V) for a distribution of 500 ratios with center and standard deviation equal to the measured line ratio and its uncertainty, respectively. H$\alpha$/H$\beta$ and H$\gamma$/H$\beta$ indicate E(B$-$V)$=$0.09$\pm$0.06 and 0.19$\pm$0.10, respectively, in statistical agreement. The weighted average is E(B$-$V)$=$0.12$\pm$0.05. We use the \textsc{PyNeb} \textsc{getCorrHb} function to perform the reddening correction relative to H$\beta$ for the emission line ratios in the G235M spectrum, and propagate the uncertainties on the relative line fluxes and E(B$-$V) to obtain uncertainties on the line intensity ratios. We report the reddening-corrected G235M line intensities relative to H$\beta$ in Table \ref{t:d40_235_linetab}.

The relative Paschen line ratios measured in the G395M spectrum indicate that a flux calibration issue is still present in the data. For example, the measured P8/P6, P7/P6, and P6/P5 ratios are all greater than the theoretical ratios at \te=1.2$\times$10$^4$ K and \den$=$10$^2$ cm$^{-3}$, the opposite trend that is expected from dust reddening. Additionally, the P8/H$\beta$ ratio agrees with the theoretical ratio, indicating negligible reddening in the NIR spectrum which is at odds with the optical Balmer line ratios. As such, we do not use the Paschen-to-H$\beta$ ratios for reddening correction in D40. However, for the following \te\ and abundance analysis we only require the strong lines of \siii. In particular, \siii$\lambda$9533 neighbors P8 at 9549 \AA, such that their relative fluxes should be insensitive to both dust attenuation and the effects of wavelength-dependent flux calibration issues. We use the theoretical P8/H$\beta$ ratio \citep[0.036 at \te=1.2$\times$10$^4$ K and \den$=$10$^2$ cm$^{-3}$,][]{stor1995} and the \siii$\lambda$9533/P8 flux ratio to obtain an estimate on \siii$\lambda$9533/H$\beta$ that is roughly corrected for reddening and flux calibration uncertainties.  The resulting line ratio is I(\siii$\lambda$9533)/I(H$\beta$) $=$ 0.29$\pm$0.03, where we have assumed a 6\% relative flux uncertainty for \siii$\lambda$9533/P8 before correcting to the H$\beta$ flux.\footnote{The direct I(\siii)/I(H$\beta$) after scaling the G395M continuum to match the G235M continuum is 0.24$\pm$0.03; adopting this intensity ratio results in a larger, but consistent, \te\siii\ than reported in \S3. Therefore, the abundance results discussed in \S4.1 are qualitatively robust to the lower intensity ratio.} This line intensity ratio is used in the derivation of \te\siii\ and S$^{2+}$/H$^+$ in the following section. We include the G395M line flux ratios relative to F(P8) in Table \ref{t:d40_235_linetab}, where we also include the inferred intensity ratios relative to H$\beta$ for lines in close proximity to P8.

\begin{deluxetable}{lcc}
\tablewidth{\textwidth}
\tabletypesize{\footnotesize}
\tablecaption{D40 Line Fluxes \& Intensities}
\tablehead{
   \colhead{} & 
   \colhead{G235M} & 
   \colhead{} \\ [-0.2cm]
   \colhead{Emission Line}  & 
   \colhead{F($\lambda$)/F(H$\beta$)} &
   \colhead{I($\lambda$)/I(H$\beta$)}}
\startdata
H$\gamma$ 4342 & 0.420 $\pm$ 0.026 & 0.451 $\pm$ 0.031 \\ 
\oiii\ 4364 & 0.081 $\pm$ 0.007 & 0.087 $\pm$ 0.008 \\ 
\ion{He}{1} 4473 & 0.039 $\pm$ 0.004 & 0.041 $\pm$ 0.004 \\ 
\ion{He}{2} 4687 & 0.016 $\pm$ 0.003 & 0.016 $\pm$ 0.003 \\ 
\ariv\ 4713 & 0.003 $\pm$ 0.003 & 0.003 $\pm$ 0.003 \\ 
\ion{He}{1} 4715 & 0.007 $\pm$ 0.003 & 0.007 $\pm$ 0.003 \\ 
H$\beta$ 4863 & 1.000 $\pm$ 0.060 & 1.000 $\pm$ 0.060 \\ 
\oiii\ 4960 & 2.014 $\pm$ 0.121 & 1.993 $\pm$ 0.120 \\ 
\oiii\ 5008 & 6.002 $\pm$ 0.361 & 5.914 $\pm$ 0.358 \\ 
\ion{He}{1} 5017 & 0.028 $\pm$ 0.004 & 0.027 $\pm$ 0.004 \\ 
\ion{He}{1} 5877 & 0.003 $\pm$ 0.003 & 0.107 $\pm$ 0.008 \\ 
\oi\ 6302 & 0.061 $\pm$ 0.004 & 0.055 $\pm$ 0.004 \\ 
\siii\ 6314 & 0.018 $\pm$ 0.002 & 0.017 $\pm$ 0.002 \\ 
\oi\ 6366 & 0.019 $\pm$ 0.002 & 0.017 $\pm$ 0.002 \\ 
\ion{Si}{2} 6373 & 0.006 $\pm$ 0.002 & 0.006 $\pm$ 0.002 \\ 
\nii\ 6550 & 0.042 $\pm$ 0.003 & 0.038 $\pm$ 0.003 \\ 
H$\alpha$ 6565 & 3.051 $\pm$ 0.183 & 2.760 $\pm$ 0.208 \\ 
\nii\ 6585 & 0.112 $\pm$ 0.007 & 0.101 $\pm$ 0.008 \\ 
\ion{He}{1} 6680 & 0.031 $\pm$ 0.002 & 0.028 $\pm$ 0.003 \\ 
\sii\ 6718 & 0.146 $\pm$ 0.009 & 0.131 $\pm$ 0.010 \\ 
\sii\ 6733 & 0.108 $\pm$ 0.007 & 0.097 $\pm$ 0.008 \\ 
\ion{He}{1} 7067 & 0.034 $\pm$ 0.003 & 0.030 $\pm$ 0.003 \\ 
\ariii\ 7138 & 0.053 $\pm$ 0.004 & 0.047 $\pm$ 0.004 \\ 
\oii\ 7322 & 0.028 $\pm$ 0.003 & 0.025 $\pm$ 0.003 \\ 
\oii\ 7332 & 0.023 $\pm$ 0.002 & 0.020 $\pm$ 0.002 \\ 
\hline 
F(H$\beta$) (erg/s/cm$^2$) & (8.65 $\pm$ 0.03)$\times$10$^{-18}$ &   \\ 
E(B$-$V) & 0.12 $\pm$ 0.05 &   \\ 
\hline 
\hline \\ 
\multicolumn{3}{c}{G395M} \\ 
Emission Line & F($\lambda$)/F(P8) & I($\lambda$)/I(H$\beta$) \\ 
\hline 
\siii\ 9071 & 3.037 $\pm$ 0.302 & 0.109 $\pm$ 0.011 \\ 
\siii\ 9533 & 7.980 $\pm$ 0.774 & 0.286 $\pm$ 0.028 \\ 
P8 9549 & 1.000 $\pm$ 0.123 & 0.036 $\pm$ 0.004 \\ 
P7 10053 & 1.899 $\pm$ 0.204 & \nodata  \\ 
\ion{He}{1} 10833 & 7.894 $\pm$ 0.769 & \nodata  \\ 
P6 10941 & 2.360 $\pm$ 0.246 & \nodata  \\ 
P5 12822 & 3.543 $\pm$ 0.352 & \nodata  \\ 
\hline 
F(P8) (erg/s/cm$^2$) & (0.32 $\pm$ 0.02)$\times$10$^{-18}$ &   \\ 
\enddata
\label{t:d40_235_linetab}
\tablecomments{Line fluxes and intensities measured from the NIRSpec rest optical and NIR data of D40. \textit{Top:} Emission lines (Left), line flux relative to H$\beta$ (Center), and reddening-corrected intensities relative to H$\beta$ (Right) measured in the G235M spectrum. The last two rows provide the flux of H$\beta$ and E(B$-$V). \textit{Bottom:} Emission lines (Left), line flux relative to P8 (Center), and intensity ratios relative to H$\beta$ using the theoretical P8/H$\beta$ ratio (Right). For the G395M data, we only report the inferred I($\lambda$)/I(H$\beta$) for lines in close proximity to P8. The last row reports the measured flux of P8.}
\end{deluxetable}


\begin{deluxetable}{lccc}  
\tablecaption{Adopted Atomic Data \label{t:atomic}}
\tablewidth{\columnwidth}
\tablehead{ 
  \colhead{Ion}	&
  \colhead{Transition Probabilities}	&
  \colhead{Collision Strengths}
  }
\startdata
N$^+$ 	& \citet{froe2004}  &  \citet{taya2011}  \\ 
O$^+$ 	& \citet{froe2004}  &  \citet{kisi2009} \\
O$^{2+}$ 	& \citet{froe2004}  &   \citet{stor2014} \\
S$^+$   & \citet{irim2005}  & \citet{taya2010} \\
S$^{2+}$    & \citet{froe2006}  &   \citet{huds2012}   \\
Ar$^{2+}$   &   \citet{mend1983}    &   \citet{muno2009}  \\
Ar$^{3+}$   &   \citet{mend1982}    &   \citet{rams1997}
\enddata
\end{deluxetable}

\section{Physical Conditions, Ionic and Total Abundances}

\subsection{Electron Temperature and Density}

To directly calculate the gas-phase chemical abundances in D40, we assume a three-zone ionization model. The low-ionization zone contains O$^+$, N$^+$, and S$^+$, with ionization potentials (IPs) 13.6 eV, 14.5 eV, and 10.4 eV, respectively; the intermediate-ionization zone contains S$^{2+}$ (IP$=$23.3 eV) and Ar$^{2+}$ (IP$=$27.6 eV); the high-ionization zone contains O$^{2+}$ (IP$=$35.1 eV) and Ar$^{3+}$ (IP$=$40.7 eV). We provide the atomic data used for \te, \den, and ionic abundance calculations in Table \ref{t:atomic}. We detect the \te-sensitive \oiii$\lambda$4364 and \siii$\lambda$6314 emission lines as well as their strong-line counterparts at high-S/N, which characterize the gas in the high- and intermediate-ionization zones, respectively. To calculate \te\ from these ions, we use the \oiii$\lambda$4364/\oiii$\lambda$5008 and \siii$\lambda$6314/\siii$\lambda$9533 ratios (where the flux of \siii$\lambda$9533 has been corrected to the H$\beta$ flux using the theoretical P8/H$\beta$ ratio, see \S2) in the \textsc{PyNeb} \textsc{getTemDen} function assuming a constant \den\ of 10$^2$ cm$^{-3}$. We use the same function and the \sii$\lambda\lambda$6718,6733 doublet to calculate \den\ at the measured \te\siii. The electron density, \den\sii$=$73$\pm$68 cm$^{-3}$, is less than the average \den\ measured in high-$z$ star-forming galaxies \citep[$\sim$250 cm$^{-3}$,][]{sand2016,stro2017} and is in the low-density limit where abundances derived from most optical CELs are relatively insensitive to \den. To determine uncertainties on \te\ and \den, we generate a distribution of 500 line ratios centered on the measured ratio and with standard deviation equal to the uncertainty on the emission line ratio. We then calculate the resulting \te\ or \den\ distribution and take the standard deviation as the uncertainty.

\begin{figure}[t]
   \centering
   \includegraphics[width=0.47\textwidth]{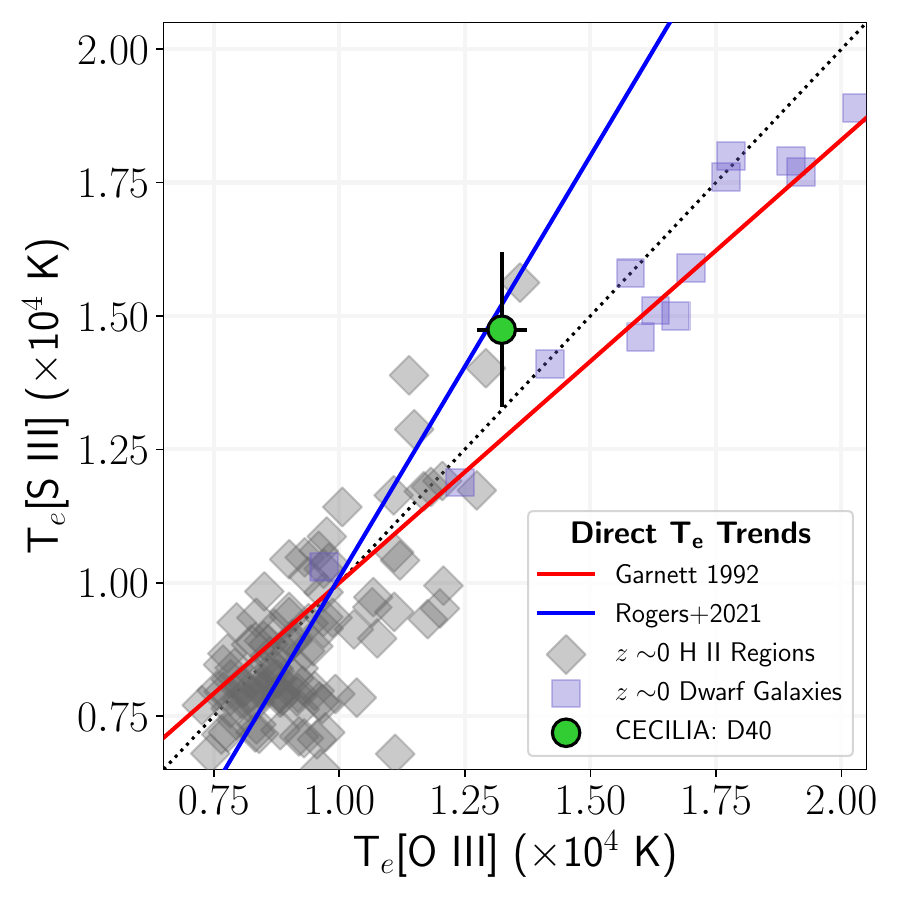}
   \caption{\te\siii-\te\oiii\ trends measured in $z$$\sim$0 extragalactic \hii\ regions (gray diamonds, see \S3) and dwarf galaxies (blue squares, see \S3), and in D40 (green circle) at $z$$\sim$3. The \citet{garn1992} photoionization model \te\siii-\te\oiii\ relation is shown as the solid red line; the empirical scaling relation of \citet{roge2021} is provided as the solid blue line. The dotted line represents equal \te\siii\ and \te\oiii. The \te\ measured in D40 agree with the \te\ trends of local ionized regions and the empirical \te-\te\ scaling relation.}
   \label{fig:ts3to3}
\end{figure}

Table \ref{t:d40_temtab} reports the physical conditions measured in D40. The direct \te\ in D40, \te\oiii$=$13200$\pm$500 K and \te\siii$=$14700$\pm$1400 K, are plotted in Figure \ref{fig:ts3to3} alongside \te\ measurements in extragalactic \hii\ regions \citep[from][]{berg2020,roge2021,roge2022} and low-metallicity dwarf galaxies \citep{berg2021,izot2021,thua2022,arel2022classy}. The empirical \te-\te\ scaling relation of \citet{roge2021} and photoionization model relation of \citet{garn1992} are provided as solid blue and red lines, respectively. The \te\oiii\ and \te\siii\ measurements in D40 are in good agreement with the \te\ measured in local star-forming systems and with the empirical \te-\te\ scaling relation, although there are few extragalactic \hii\ regions with similarly high \te\oiii\ and \te\siii. While this relation appears to characterize the \te\ in D40, we emphasize that large sample sizes are required to assess the scatter about a given \te-\te\ relation \citep[e.g.,][]{arel2020,yate2020,roge2021} and whether these relations describe the bulk \te\ trends at high-$z$.

While we detect intense \oii$\lambda\lambda$7322,7332 auroral line emission in D40, we lack the nebular lines to estimate the low-ionization zone \te\ directly from the NIRSpec data. Future work will utilize ground-based \oii$\lambda\lambda$3727,3729 detections to directly measure \te\oii\ in the CECILIA galaxies. Given the good agreement with the empirical \te\oiii-\te\siii\ relation, we infer the low-ionization zone \te\ using the weighted average approach recommended by \citet{roge2021}. First, we use the direct \te\oiii\ and \te\siii\ to estimate the low-ionization zone temperature using Equations 3 and 6 in \citet{roge2021}. We also apply their Equation 8 to account for the intrinsic \te\ scatter about each relation when calculating the uncertainty on the inferred \te. Next, we take the weighted average of the two inferred \te\ and adopt this as T$_{e,Low}$. Finally, we use the largest inferred \te\ uncertainty as the uncertainty on T$_{e,Low}$. The inferred low-ionization zone temperature, T$_{e,Low}$=12300$\pm$1000 K, is also reported in Table \ref{t:d40_temtab}.

\subsection{Abundances and Ionization Correction Factors}

We calculate ionic abundances relative to H$^+$ using the \textsc{PyNeb getIonAbundance} function, the strong-line intensities, the ionization zone \te, and \den\sii. Uncertainty on the line intensity and emissivity ratios (the latter of which is dependent on the \te\ uncertainty) are added in quadrature to calculate the ionic abundance uncertainty. Table \ref{t:d40_temtab} reports the direct and inferred ionic abundances measured in D40. The inferred O$^+$/H$^+$ abundance is calculated from the \oii\ auroral lines, which are sensitive to \te\ and \den. While some studies have found good consistency between O$^+$/H$^+$ abundances derived using the nebular and auroral \oii\ lines \citep{knia2003,knia2004}, others with direct low- and high-ionization zone \te\ measure inconsistent O$^+$/H$^+$ abundances from the different \oii\ lines \citep{rodr2020}. Furthermore, \citet{mend2023} recently showed that \den\ fluctuations have the potential to bias O$^+$/H$^+$ abundances upwards by $\sim$0.1 dex when calculated using the \oii\ auroral lines. Such fluctuations are not explicitly accounted for here, but the relatively large O$^+$/H$^+$ uncertainty in D40 reflects the \te-sensitivity of \oii$\lambda\lambda$7322,7332 auroral lines when used for ionic abundances. O$^+$ and O$^{2+}$ are the two relevant ionic states of O in the ISM for systems ionized by O and B stars \citep{berg2021}. While we detect intense [\ion{O}{1}]$\lambda$6302 emission, this emission could be reflective of the photodissociation region rather than the ionized gas. As such, we assume the total O abundance is described by O/H $\approx$ (O$^+$ + O$^{2+}$)/H$^+$. The resulting O/H abundance in D40 is 12+log(O/H) $ = 8.07\pm0.06$ dex, $\sim$24\% of the solar abundance (see \S4). We do not correct for dust grain depletion of O, which is most significant in high-metallicity environments \citep[up to 0.1 dex at solar metallicity, see][]{peim2010,pena2012}. Given the similar IP of N$^+$ and O$^+$, we adopt the usual approximation that N/O $\approx$ N$^+$/O$^+$, which is good to within 10\% of the true gas-phase N/O \citep[see][]{nava2006,amay2021}.

\begin{deluxetable}{lc}
\tablewidth{\textwidth}
\tabletypesize{\footnotesize}
\tablecaption{D40 Physical Conditions}
\tablehead{
   \colhead{Property}  & 
   \colhead{Measurement}}
\startdata
\te\oiii\ (K) & 13200 $\pm$ 500 \\ 
\te\siii\ (K) & 14700 $\pm$ 1400 \\ 
T$_{e,Low}$  (K, inferred) & 12300 $\pm$ 1000 \\ 
\den\sii\ (cm$^{-3}$) & 73 $\pm$ 71 \\ 
  &   \\ 
O$^+$/H$^+$ ($\times$10$^5$) & 3.0 $\pm$ 1.3 \\ 
O$^{2+}$/H$^+$ ($\times$10$^5$) & 8.8 $\pm$ 1.1 \\ 
12+log(O/H) (dex) & 8.07 $\pm$ 0.06 \\ 
  &   \\ 
N$^+$/H$^+$ ($\times$10$^6$) & 1.3 $\pm$ 0.3 \\ 
log(N/O) (dex) & $-$1.37 $\pm$ 0.21 \\ 
  &   \\ 
S$^+$/H$^+$ ($\times$10$^7$) & 3.5 $\pm$ 0.7 \\ 
S$^{2+}$/H$^+$ ($\times$10$^7$) & 8.9 $\pm$ 1.7 \\ 
S ICF & 1.24 $\pm$ 0.12 \\ 
12+log(S/H) (dex) & 6.19 $\pm$ 0.08 \\ 
log(S/O) (dex) & $-$1.88 $\pm$ 0.10 \\ 
  &   \\ 
Ar$^{2+}$/H$^+$ ($\times$10$^7$) & 1.7 $\pm$ 0.4 \\ 
Ar ICF & 1.07 $\pm$ 0.11 \\ 
12+log(Ar/H) (dex) & 5.27 $\pm$ 0.10 \\ 
log(Ar/O) (dex) & $-$2.80 $\pm$ 0.12 \\ 
  &   \\ 
Ar$^{3+}$/H$^+$ ($\times$10$^7$)$_u$ & 1.0 $\pm$ 0.5 \\ 
Ar ICF$_u$ & 1.01 $\pm$ 0.10 \\ 
log(Ar/O)$_u$ (dex) & $<$$-$2.63 \\ 
\enddata
\label{t:d40_temtab}
\tablecomments{\te, \den, ionic abundances, ICFs, and total abundances measured in D40. We also report (Ar$^{3+}$/H$^+$)$_u$, the upper limit on Ar$^{3+}$ abundance assuming the combined \ariv$\lambda$4713 and \ion{He}{1}$\lambda$4715 emission is entirely \ariv. The resulting change to the ICF, Ar ICF$_u$, and relative abundance, log(Ar/O)$_u$, are reported in the last two rows.}
\end{deluxetable}

For other elements, ionization correction factors (ICFs) are required to account for the abundances of unobserved ionic species. Given an ionic abundance X$^{i+}$/H$^+$, the total elemental abundance X/H is calculated as X/H = ICF(X)$\times$X$^{i+}$/H$^+$ where ICF(X) depends on the ionization state of the gas. For example, there are no observable [\ion{S}{4}] emission lines in the optical or NIR, but the missing S$^{3+}$ must be accounted for when calculating total S abundances. Similarly, correction for Ar$^+$ is required for total Ar abundance, as is an additional correction for Ar$^{3+}$ in very high-ionization systems if the optical \ariv$\lambda\lambda$4713,4742 lines are not detected (see the discussion at the end of this section).

Owing to the challenges associated with measuring empirical total abundances, ICFs are calibrated via two methods: with similarities in IP \citep[like N and Ne,][]{peim1969}, and with photoionization models \citep{thua1995,izot2006,amay2021}. The latter are required for S and Ar abundances, but these photoionization models necessarily make assumptions about the ionizing spectrum that may not describe high-$z$ galaxies, which have generally harder ionizing spectra. For example, \citet{amay2021} developed ICFs from the Bayesian Oxygen and Nitrogen Determinations \citep[][]{vale2016} grid of models, which utilize PopStar single-star SEDs \citep{moll2009} with varying gas-phase O/H and N/O. To calibrate these ICFs, \citet{amay2021} weighted each model based on its ability to reproduce the emission line trends of local blue compact galaxies and \hii\ regions on the classic Baldwin-Philips-Terlevich (BPT) diagram \citep{bald1981}, or log(\oiii$\lambda$5008/H$\beta$) vs.\ log(\nii$\lambda$6585/H$\alpha$). It is well-established that high-$z$ star-forming galaxies are offset from the $z$$\sim$0 BPT star-forming locus \citep{mast2014,stei2014,shap2015,stro2017,sand2023c}, and it follows that these ICFs may not be suitable for high-$z$ galaxies or other sources with harder ionizing spectra. While some studies have considered harder ionizing sources when calibrating ICFs \citep[e.g.,][]{berg2021}, it is unclear if these can reproduce the general emission line trends of high-$z$ galaxies.

For the present study, we use the S and Ar ICFs of \citet{izot2006} for abundances in D40. These ICFs are dependent on O$^+$/O and have different functional forms depending on 12+log(O/H); as recommended by \citet{izot2006}, we use a linear interpolation of intermediate- and high-metallicity ICFs for D40. However, we emphasize that the lack of available ICFs that are calibrated either on the strong-line trends in high-$z$ galaxies or model grids which consider other ionizing sources \citep[e.g., binary stars,][]{eldr2017,stan2018} presents one of the largest uncertainties when attempting to estimate the total abundance of elements like S and Ar. To partially account for this, we assume a 10\% uncertainty in applying the \citet{izot2006} S and Ar ICFs. In future works, we will leverage the CECILIA sample and new grids of photoionization models to derive physically-motivated ICFs for reliable total abundance inferences in high-$z$ environments.

Finally, we note that the blend of [\ion{Ar}{4}]$\lambda$4713 and \ion{He}{1}$\lambda$4715 is detected in D40. In high-ionization sources, the ionic fraction of Ar$^{3+}$ becomes comparable to that of Ar$^{2+}$ \citep[e.g.,][]{berg2021}; as such, [\ion{Ar}{4}] emission permits a direct measurement of Ar$^{3+}$/H$^+$ and a total Ar/H that is less susceptible to the uncertainties of the adopted ICF (i.e., the ICF is only accounting for the missing Ar$^+$/H$^+$ abundance). However, the Gaussian fit to [\ion{Ar}{4}]$\lambda$4713 indicates that this line is only weakly detected (S/N $\sim$ 1). As an exercise meant to provide an upper limit on Ar/H, we attribute the [\ion{Ar}{4}]$\lambda$4713 and \ion{He}{1}$\lambda$4715 blend entirely to [\ion{Ar}{4}], calculate the resulting Ar$^{3+}$/H$^+$ abundance using \te\oiii, and determine Ar/H using (Ar$^{2+}$ + Ar$^{3+}$)/H$^+$ and the alternative Ar ICF proposed by \citet{izot2006}. The resulting upper limit on Ar/O is reported as log(Ar/O)$_u$ in Table \ref{t:d40_temtab}.

\begin{figure}[ht]
   \centering
   \includegraphics[height=0.80\textheight]{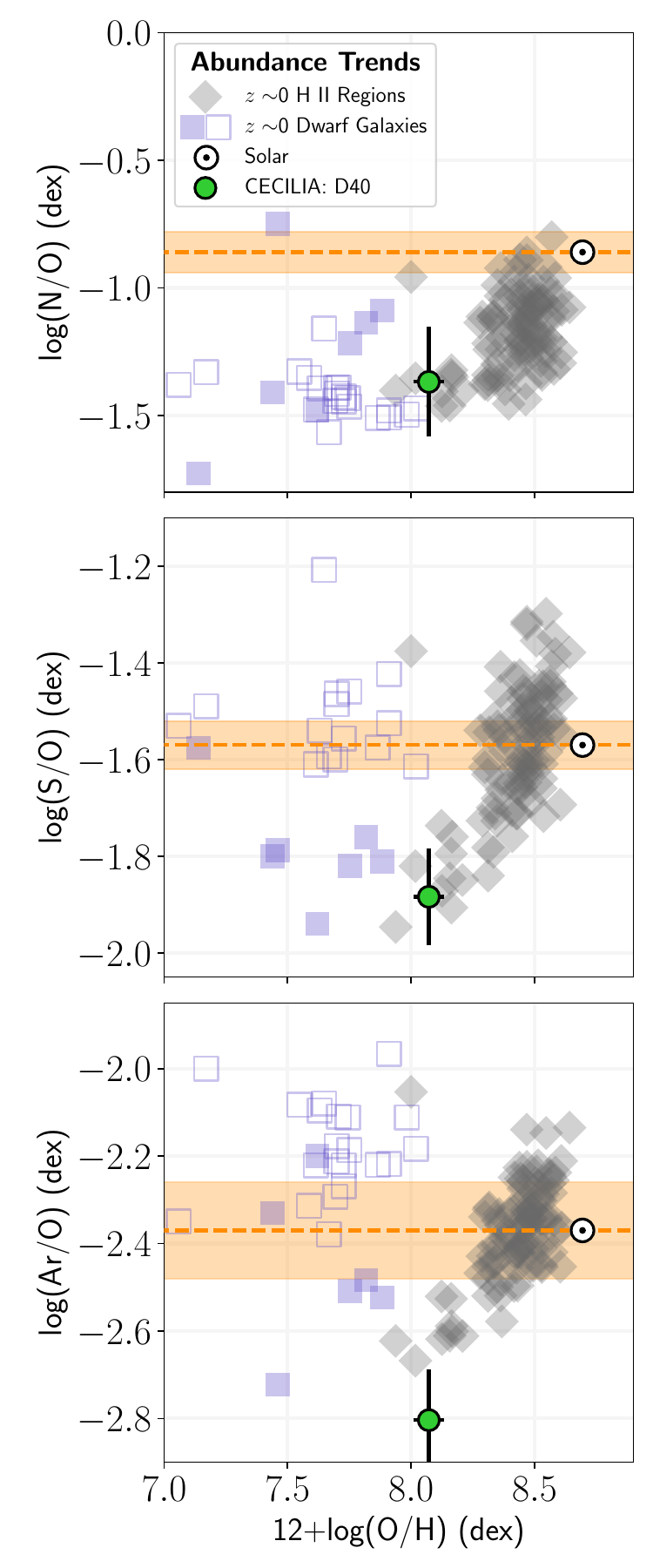}
   \caption{Abundance trends of D40 (green circle), local \hii\ regions (gray diamonds), and low-metallicity dwarf galaxies (blue squares). Open blue squares represent dwarf galaxies without direct \te\siii. The $\odot$ denotes solar abundances from \citet{aspl2021}; the dashed orange line and shaded areas are the solar N/O, S/O, and Ar/O ratios. \textit{Top Panel:} log(N/O) vs.\ 12+log(O/H). \textit{Middle Panel:} log(S/O) vs.\ 12+log(O/H). \textit{Bottom Panel:} log(Ar/O) vs.\ 12+log(O/H). The first \te-based S and Ar abundances at $z$$\sim$3 indicate sub-solar S/O and Ar/O, suggesting that these three elements are sensitive to O/H or star-formation history.}
   \label{fig:abun}
\end{figure} 

\section{Discussion}

\subsection{Abundance Results}

We now compare the abundance trends measured in D40 to local star-forming systems, illustrated in Figure \ref{fig:abun}. The relative abundances of N/O, S/O, or Ar/O are plotted against the gas-phase O/H abundance. The local \hii\ regions and dwarf galaxies from Figure \ref{fig:ts3to3} (i.e., those with simultaneous \te\oiii\ and \te\siii) are represented as gray diamonds and blue squares, respectively. We denote abundances in $z$$\sim$0 dwarf galaxies from \citet{skil2013}, \citet{berg2019}, and \citet{aver2022} that have measurements of \te\oiii\ but are missing \te\siii\ as open blue squares. We also represent solar abundances reported in \citet{aspl2021} as the $\odot$ symbol and dotted orange lines (fixed solar O/H), with the shaded regions indicating the uncertainty.

The O/H abundance measured in D40, 12+log(O/H) $ = 8.07\pm0.06$ dex, is indicative of a relatively metal-poor ISM. While this is lower than the average model-inferred metallicity of the KBSS galaxies reported in \citet{stro2018}, D40's O/H is consistent with its relatively high \te\oiii\ and \te\siii\ and is in general agreement with emerging direct O/H abundance trends at $z$$>$1 \citep[e.g.,][]{sand2023,welc2024}. In the top panel, the inferred log(N/O) abundance in D40 (green circle, $-$1.37$\pm$0.21 dex) is in good agreement with the primary N/O plateau \citep[log(N/O) $\sim$ $-$1.43 dex,][]{garn1990,vanZ2006,nava2006} established by high-ionization, low-metallicity dwarf galaxies and \hii\ regions. The metallicity of D40 is close to the 12+log(O/H) value where secondary N production from intermediate-mass stars is expected to become significant \citep[12+log(O/H) $\sim$ 8.3 dex, see][]{henr2000}, but given that we do not observe enhanced N/O, the N enrichment in D40 likely arises from primary nucleosynthesis from massive stars. We note that the large uncertainty on the log(N/O) relative abundance is related to the dependence of N$^+$ and O$^+$ on the inferred low-ionization zone \te\ and the use of the \oii\ auroral lines for O$^+$ abundance.

O, S, and Ar are all generally thought to be primarily produced by the $\alpha$ process in massive stars and are released to the ISM on relatively short timescales via CCSN. In this scenario, the relative abundances of S/O and Ar/O are related to CCSN yields and independent of metallicity. Therefore, it is of interest to assess any potential metallicity or redshift evolution of relative $\alpha$ abundances, which would suggest alternative enrichment mechanisms for the different elements. The inferred gas-phase S/O abundance measured in D40, log(S/O) $ =-1.88\pm0.10$ dex, is 2$\sigma$ below the solar abundance ratio\footnote{We emphasize that the reported uncertainties on S/O and Ar/O do not reflect systematic errors related to the ICFs, which may have different functional forms in high-$z$ galaxies (see discussion in \S3.2).} but is consistent with the lowest S/O abundances reported in local ionized regions. The average S/O observed in the local objects, $\langle$log(S/O)$_{z=0}\rangle =-1.58\pm0.14$ dex, shows relatively large dispersion, which might be related to the challenges associated with observing the \siii$\lambda\lambda$9071,9533 lines from ground-based observatories. For example, contamination from telluric absorption features \citep{noll2012} can bias the NIR \siii\ emission line intensities low and the derived \te\siii\ high, resulting in large dispersion in S/H and S/O. Telluric contamination, however, does not affect our JWST/NIRSpec observations of D40's \siii\ lines or S/O abundance, the latter of which is consistent with the S/O of \hii\ regions with similarly low 12+log(O/H). Taken together with the S/O abundances in dwarf galaxies with measured \te\siii, there is tentative evidence for a more complex enrichment mechanism for S that leads to deviations from the abundance patterns expected from CCSN enrichment alone.

The bottom panel of Figure \ref{fig:abun} reveals that the inferred relative Ar/O abundance in D40, log(Ar/O) $ =-2.80\pm0.12$ dex, is also significantly sub-solar. The same conclusion is reached when considering the upper limit of Ar/O from the maximum [\ion{Ar}{4}] intensity, log(Ar/O)$_u <-2.63$ dex. Other available Ar ICFs produce  similarly low Ar/O, with the maximum log(Ar/O) being $-$2.60 dex when using the \citet{thua1995} ICFs. Furthermore, accounting for the depletion of metals onto dust grains would increase the O abundance and only further decrease Ar/O. The average log(Ar/O) for the $z$$\sim$0 SF systems is $\langle$log(Ar/O)$_{z=0}\rangle =-2.38\pm0.15$ dex, in agreement with the \citet{aspl2021} solar ratio. Both $\langle$log(Ar/O)$_{z=0}\rangle$ and the solar Ar/O are larger than the gas-phase log(Ar/O) in D40 by $>$2$\sigma$.

To further explore these abundance trends, we investigate the relative Ar/S abundance in D40. The log(Ar/S) abundance measured when using the inferred total S/H and Ar/H abundances reported in Table \ref{t:d40_temtab} is log(Ar/S) $= -0.92\pm0.13$ dex. While this is below the solar abundance ratio of log(Ar/S)$_\odot = -0.74\pm0.10$ dex, it is also sensitive to the shortcomings in the adopted S and Ar ICFs. For example, if we instead consider the maximum Ar abundance from the upper limit on Ar$^{3+}$/H$^+$, we find log(Ar/S)$_u < -0.76$ dex, which agrees with the solar ratio. Even this upper limit is sensitive to the aforementioned shortcomings of the ICFs, which may fail to account for a significant fraction of S$^{3+}$ and Ar$^+$ in the ISM. Given the similar IPs of S$^+$ (23.3 eV) and Ar$^+$ (27.6 eV), log(Ar$^{2+}$/S$^{2+}$) has been proposed as an alternative, ICF-independent measure of log(Ar/S) \citep{kenn2003b,crox2016}. For D40, we measure a direct log(Ar$^{2+}$/S$^{2+}$) $ =-0.71\pm0.12$ dex, in good agreement with the solar log(Ar/S) ratio. Taken together with the abundance trends from Figure \ref{fig:abun}, the ISM in D40 at $z$$\sim$3 is S and Ar deficient relative to O when compared to the gas-phase abundances in the local universe, yet the relative Ar/S abundance at high-$z$ and in local objects is consistent with a joint evolution of Ar and S.

\subsection{S/O, Ar/O, and Implications for \texorpdfstring{$\alpha$}{a} Enhancement}


While it is often assumed that the relative $\alpha$ element abundance ratios are independent of 12+log(O/H), ionization, or star-formation history, recent nucleosynthesis models have revealed that Type Ia SNe can produce certain $\alpha$ elements. For example, the models of \citet{koba2020b,koba2020a} indicate that the relative yields of $^{32}$S/$^{56}$Fe and $^{36}$Ar/$^{56}$Fe from different Type Ia progenitors can match the observed solar S/Fe and Ar/Fe abundance ratios, respectively. Furthermore, these models indicate that the Type Ia SNe yield of S and Ar are roughly the same and significantly greater than the yield of O in different Type Ia progenitors \citep[see][]{koba2020a}. This suggests that the relative abundances of $\alpha$ elements like S/O and Ar/O trace enrichment by both CCSN and Type Ia SNe in a manner similar to Fe/O. Indeed, galactic chemical evolution models can match the Ar/O and Ar/H abundances of stars and planetary nebulae in M31 using various star-formation histories to account for the different mechanisms of Ar production \citep{arna2022,koba2023}.

In this scenario, rapid enrichment from CCSN results in enhanced O relative to S and Ar, producing lower gas-phase log(S/O) and log(Ar/O) abundance ratios. Type Ia SNe enrich the ISM with additional S and Ar, thereby increasing log(S/O) and log(Ar/O) after a large time delay \citep[from $\sim$0.1 to $>$ 10 Gyr,][]{koba2009,koba2020a}. Unlike the log(S/O) and log(Ar/O) abundance ratios, the log(Ar/S) abundance should remain relatively constant \citep[consistent with direct abundance trends in the local universe,][]{stev1993} if the enrichment of these elements is set by CCSN at early times and if the yield of S and Ar are similar for Type Ia SNe at late times \citep{koba2020a}. The lifetime of the massive stars ionizing the ISM is much shorter than the Type Ia SNe time delay, such that the chemical compositions of the \hii\ regions are reflective of the stars that were born in this environment. Considering this time delay and the age of the Universe at D40's redshift ($\sim$2 Gyr for $z$$=$2.9628), the S/O and Ar/O abundance trends in D40 could be explained by predominant enrichment from CCSN that results in enhanced gas-phase O relative to S and Ar.

The S and Ar abundance trends presented here are the first to be directly calculated from a measured \te\siii\ and strong lines of \siii\ and \ariii\ in a galaxy at $z$$>$0.2. Currently, no prior observations have constrained \te\siii\ in high-$z$ galaxies, which is necessary for reliable S/O and Ar/O abundances. Additionally, the \siii\ and \ariii\ strong lines require both broad wavelength coverage and fairly deep integration times to detect. For example, \citet{isob2023} only detect \ariii$\lambda$7138 in two galaxies at $z$$>$4, but \te\oiii\ and \te\siii\ are unconstrained in both galaxies and all other Ar abundances are upper limits. Furthermore, the S/O abundances from \citet{isob2023} are determined using the \sii\ emission lines alone, which accounts for only a small fraction of the total S abundance in high-ionization galaxies. For example, the S$^{2+}$/H$^+$ abundance in D40 is $>$2.5 times that of S$^{+}$/H$^+$ (see Table \ref{t:d40_temtab}), implying that a measure of only S$^{+}$/H$^+$ is heavily reliant on the validity of the S ICF applied (which we note is one of the largest uncertainties related to high-$z$ abundance inferences, see discussion in \S3.2).

If these first constraints on the gas-phase S/O and Ar/O abundance trends at high-$z$ are the result of primary enrichment from CCSN, then these data add to the growing observational evidence of a disconnect between the nucleosynthesis products of Type II and Ia SNe in the early universe. As mentioned in \S3.2, the emission line ratios of high-$z$ star-forming galaxies are offset in the BPT diagram, a trend that can be explained with a harder ionizing spectrum at fixed O/H \citep{stei2014,stro2017,sand2020,topp2020,runc2021}. By decoupling the stellar and nebular metallicities in photoionizaiton models, \citet{stro2018} demonstrated that the bulk emission line trends of $z$$>$2 galaxies can be reproduced with lower stellar metallicity (to reduce the line-blanketing by Fe lines) relative to the nebular metallicity (which influences the strong optical emission lines such as \oiii$\lambda$5008). Stacks of rest-UV spectra of high-$z$ star-forming galaxies arrive at similar conclusions: spectral synthesis of low-metallicity stellar populations is required to match the UV features which are dominated by Fe, but larger gas-phase O abundance is needed to produce the observed strong line ratios \citep[e.g.,][]{stei2016,cull2019,topp2020}. A pattern of sub-solar S/O and Ar/O relative abundances with simultaneous solar Ar/S abundance ratios at high-$z$ would suggest that the differences in enrichment timescales play a crucial role in the gas-phase conditions observed in early star-forming galaxies. Utilizing S/O and Ar/O as tracers of Type Ia SNe and O/H as a tracer of CCSN enrichment has the additional benefit that the required CELs are entirely observable in the rest-frame optical/NIR. Therefore, the \ariii$\lambda$7138 emission line may act as an accessible tracer of these different enrichment mechanisms, providing further insight into chemical enrichment histories of high-$z$ galaxies.

\section{Conclusions}

We report the chemical abundance trends in Q2343-D40, a galaxy at $z$$=$2.9628$\pm$0.0001 observed with JWST/NIRSpec as part of the CECILIA program. For the first time in a galaxy at $z$$>$0.2, we calculate \te\siii\ from the \siii$\lambda$6314 auroral line and the \siii$\lambda$9533 strong line. We utilize the high-S/N detections of \oiii$\lambda$4364 and \oiii$\lambda$5008 to directly calculate the high-ionization zone \te. The \te\ trends in D40, \te\oiii$=$13200$\pm$500 K and \te\siii$=$14700$\pm$1400 K, are in good agreement with the \citet{roge2021} empirical \te-\te\ relation (see Figure \ref{fig:ts3to3}), but we note that other surveys have found large scatter in \te\siii\ at fixed \te\oiii. While \te\siii\ is challenging to measure, owing to the broad wavelength coverage required to measure \siii$\lambda\lambda$9071,9533 and the low intensity of the \siii\ auroral line, simultaneous measurements of \te\siii\ and \te\oiii\ are required to assess whether \te-\te\ scaling relations calibrated on local ionized nebulae or photoionization models are appropriate to apply at high-$z$.

We use these direct \te\oiii\ and \te\siii\ values to measure the gas-phase ionic abundances in D40, representing the first direct abundances of S$^{2+}$/H$^+$ and Ar$^{2+}$/H$^+$ in a high-$z$ galaxy. Using an inferred low-ionization zone \te\ and the necessary ICFs, we calculate the inferred total O/H abundance and the relative abundances of N/O, S/O, and Ar/O (plotted in Figure \ref{fig:abun}). The O/H abundance in D40, 12+log(O/H) $=8.07\pm0.06$, is $\sim$24\% solar, and the N/O abundance agrees with the primary N/O plateau observed in similarly metal-poor galaxies in the local universe. The relative S/O abundance in D40 is below the solar abundance ratio by more than 2$\sigma$ (0.31 dex), but is in agreement with the lowest S/O abundances observed in local star-forming regions. The S/O abundance dispersion in local ionized regions is possibly related to the observational challenges associated with measuring the NIR \siii\ lines from the ground. Alternatively, the scatter in S/O may be related to the ICFs, which must account for a large fraction of unobserved S$^{3+}$ in regions characterized by high-ionization.

The Ar/O abundance in D40, log(Ar/O) $ =-2.80\pm0.12$ dex, is $>$2$\sigma$ (0.43 dex) below the solar abundance ratio and the average log(Ar/O) observed in $z$$\sim$0 star-forming systems. This result is robust to the inclusion of Ar$^{3+}$ from the maximum possible [\ion{Ar}{4}]$\lambda$4713 intensity, reddening correction uncertainties, and the effects of dust depletion. While the inferred S/O abundance is sensitive to the functional form of the ICF, a systematic error with the ICF is less likely to alter the estimated maximum Ar/O because such a scenario would imply a significant fraction of missing Ar$^+$ in the high-ionization ISM of D40. Comparing the direct Ar$^{2+}$ and S$^{2+}$ abundances in D40, we measure log(Ar$^{2+}$/S$^{2+}$) $\approx$ log(Ar/S) $=-0.71\pm0.12$ dex, in good agreement with the solar Ar/S abundance.

Modern nucleosynthesis models find that S and Ar are produced via the $\alpha$ process in massive stars and in Type Ia SNe \citep{koba2020b,koba2020a}. In this way, S/O and Ar/O are sensitive to the relative contributions from Type II and Ia SNe. Additionally, the yield of S and Ar are similar for different Type Ia progenitor models, implying that the Ar/S ratio remains relatively constant from both CCSN and Type Ia enrichment. As such, the inferred sub-solar S/O and Ar/O abundances, coupled with the Ar/S ratio that matches the solar value, could be the result of primarily CCSN enrichment with small contributions from Type Ia SNe, an indirect probe of the relative enrichment of O with respect to Fe.

If future JWST observations and more secure ICFs confirm that low S/O and Ar/O abundances are common in high-$z$ galaxies, then such abundance trends would corroborate the growing evidence of enhanced O relative to the nucleosynthesis products of Type Ia SNe, including the hard ionizing spectra required to reproduce the BPT offset and the inferred stellar abundances measured from stacked rest-frame UV spectra of high-$z$ galaxies. Abundances like N/O, S/O, and Ar/O provide another route to explore these relative enrichment pathways and the physical conditions within these systems. With deep rest-frame UV, optical, and NIR spectra of many $z$$>$2 star-forming galaxies, the CECILIA sample is poised to address these relative abundance trends in the early universe where the timescales of Type II and Ia SNe enrichment play an important role.

\begin{acknowledgments}
We are grateful to the referee for their constructive review of the manuscript and their insightful recommendations.
N.S.J.R. is supported under JWST-GO-02593, which was provided by NASA through a grant from the Space Telescope Science Institute, which is operated by the Association of Universities for Research in Astronomy, Inc., under NASA contract NAS 5-03127. A.L.S., G.C.R., and R.F.T. acknowledge partial support from the JWST-GO-02593.008-A, JWST-GO-02593.004-A, and JWST-GO-02593.006-A grants, respectively. R.F.T. also acknowledges support from the Pittsburgh Foundation (grant ID UN2021-121482) and the Research Corporation for Scientific Advancement (Cottrell Scholar Award, grant ID 28289).

This work is primarily based on observations made with NASA/ESA/CSA JWST, associated with PID 2593, which can be accessed via doi:\dataset[10.17909/x66z-p144]{https://doi.org/10.17909/x66z-p144}. The data were obtained from the Mikulski Archive for Space Telescopes (MAST) at the Space Telescope Science Institute, which is operated by the Association of Universities for Research in Astronomy, Inc., under NASA contract NAS 5-03127 for JWST.

\end{acknowledgments}

\textit{Facilities:} JWST (NIRSpec)


\textit{Software:} \textsc{matplotlib} \citep{hunt2007}, \textsc{NumPy} \citep{harr2020}, \textsc{PyNeb} \citep{luri2012,luri2015L}, \textsc{SciPy} \citep{virt2020}, \textsc{calwebb} \citep{calwebb_v1.12.5_2023}, \textsc{grizli} \citep{griz2023}, \textsc{msaexp} \citep{msaexp2022}, \textsc{NSClean} \citep{raus2023}.

\bibliographystyle{aasjournal}
\bibliography{aggregate_refs}

\end{document}